\documentclass[%
 reprint,
 amsmath,amssymb,
 aps,
]{revtex4-2}

\usepackage{soul}
\usepackage{siunitx}
\usepackage{booktabs}
\usepackage{graphicx}
\usepackage{dcolumn}
\usepackage{bm}
\usepackage[mathlines]{lineno}
\newcommand\numberthis{\addtocounter{equation}{1}\tag{\theequation}}

\begin{document}


\title{Search for Axion Dark Matter from  1.1 to 1.3 GHz with ADMX}

\author{G. Carosi}
\author{C. Cisneros}
\author{N. Du}
  \altaffiliation{du4@llnl.gov }
\author{S. Durham}
\author{N. Robertson}
\affiliation{Lawrence Livermore National Laboratory, Livermore, California 94550, USA}

\author{C. Goodman}
\author{M. Guzzetti}
\author{C. Hanretty}
\author{K. Enzian}
 \author{L. J Rosenberg}
   \author{G. Rybka}
\author{J. Sinnis}
  \author{D. Zhang}
  \affiliation{University of Washington, Seattle, Washington 98195, USA}
   
\author{John Clarke}
\author{I. Siddiqi}
  \affiliation{University of California, Berkeley, California 94720, USA}
\author{A. S. Chou} 
 \author{M. Hollister}
\author{A. Sonnenschein} 
  \affiliation{Fermi National Accelerator Laboratory, Batavia, Illinois 60510, USA}

\author{S. Knirck}
\affiliation{Harvard University, Cambridge, Massachusetts 02138, USA}

\author{T. J. Caligiure}
\author{J. R. Gleason}
\author{A. T. Hipp}
\author{P. Sikivie}
\author{M. E. Solano}
\author{N. S. Sullivan}
\author{D. B. Tanner}
\affiliation{University of Florida, Gainesville, Florida 32611, USA}

\author{R. Khatiwada}
\affiliation{Illinois Institute of Technology, Chicago, Illinois 60616, USA}
\affiliation{Fermi National Accelerator Laboratory, Batavia, Illinois 60510, USA}

\author{L. D. Duffy}
  \affiliation{Los Alamos National Laboratory, Los Alamos, New Mexico 87545, USA}

\author{C. Boutan}
\author{T. Braine}
\author{E. Lentz}
\author{N. S. Oblath}
\author{M. S. Taubman}
  \affiliation{Pacific Northwest National Laboratory, Richland, Washington 99354, USA}

\author{E. J. Daw}
\author{C. Mostyn}
  \author{M. G. Perry}
  \affiliation{University of Sheffield, Sheffield S10 2TN, UK}

\author{C. Bartram}
\author{J. Laurel}
\author{A. Yi}
\affiliation{SLAC National Accelerator Laboratory, 2575 Sand Hill Road, Menlo Park, California 94025, USA}

\author{T. A. Dyson}
\author{S. Ruppert}
\author{M. O. Withers}
\affiliation{Stanford University, Stanford, CA 94305, USA}

\author{C. L. Kuo}
\affiliation{SLAC National Accelerator Laboratory, 2575 Sand Hill Road, Menlo Park, California 94025, USA}
\affiliation{Stanford University, Stanford, CA 94305, USA}

\author{B. T. McAllister}
\affiliation{Swinburne University of Technology, John St, Hawthorn VIC 3122, Australia}

\author{J. H. Buckley}
\author{C. Gaikwad}
\author{J. Hoffman}
\author{K. Murch}
  \affiliation{Washington University, St. Louis, Missouri 63130, USA}
  \author{M. Goryachev}
  \author{E. Hartman}
\author{A. Quiskamp}
\author{M. E. Tobar}
\affiliation{University of Western Australia, Perth, Western Australia 6009, Australia}

\collaboration{ADMX Collaboration}

\date{\today}

\begin{abstract}
Axion dark matter can satisfy the conditions needed to account for all of the dark matter and solve the strong \textit{CP} problem. The Axion Dark Matter eXperiment (ADMX) is a direct dark matter search using a haloscope to convert axions to photons in an external magnetic field. Key to this conversion is the use of a microwave resonator that enhances the sensitivity at the frequency of interest. The ADMX experiment boosts its sensitivity using a dilution refrigerator and near quantum-limited amplifier to reduce the noise level in the experimental apparatus. In the most recent run, ADMX searched for axions between 1.10-1.31 GHz to extended Kim-Shifman-Vainshtein-Zakharov (KSVZ) sensitivity. This Letter reports on the results of that run, as well as unique aspects of this experimental setup.
\end{abstract}

\maketitle


This Letter reports the results of a search by the Axion Dark Matter eXperiment (ADMX) covering the frequency range from 1.10-1.31 GHz (4.54-5.41 $\si{\micro{eV}}$), over which we achieved sensitivity to plausible models for the quantum chromodynamic (QCD) axion. Our new results describe a previously unexplored region of parameter space for axion dark matter.

The axion is a hypothetical particle that emerges from the Peccei-Quinn (PQ) solution to the strong \textit{CP} problem \cite{Peccei1977June,Weinberg:1977ma,Wilczek:1977pj}.  In addition to solving the strong \textit{CP} problem, axions may be produced abundantly in the early universe through the misalignment mechanism, allowing them to account for all the dark matter in the universe~\cite{PRESKILL1983127,ABBOTT1983133,DINE1983137,PhysRevLett.50.925}. In the case where PQ symmetry was broken before cosmological inflation, axions over a broad mass range can elegantly account for all the dark matter in the universe~\cite{Borsanyi2016,PhysRevD.96.095001}. If PQ was broken after inflation, theory suggests that the axion mass lies in the $\mathcal{O}(1\text{--}100)~\mu$eV range~\cite{PhysRevD.83.123531,PhysRevD.91.065014,PhysRevD.92.034507,Fleury_2016,Bonati2016,PETRECZKY2016498,Borsanyi2016,PhysRevLett.118.071802,Klaer_2017,PhysRevD.96.095001,PhysRevLett.124.161103,10.21468/SciPostPhys.10.2.050,buschmann2021dark}. The coupling of axions to photons is model dependent, but two benchmark models are commonly used: the Kim-Shifman-Vainshtein-Zakharov (KSVZ)  axion \cite{Kim:1979if,Shifman:1979if} and the more feebly coupled Dine-Fischler-Srednicki-Zhitnitsky (DFSZ) axion \cite{Dine:1981rt,Zhitnitsky:1980tq}. 

ADMX searches for dark matter axions using an axion haloscope~\cite{PhysRevLett.51.1415,RevModPhys.82.557,RevModPhys.93.015004}, which uses the Primakoff effect resonantly to convert axions to photons in a strong magnetic field. The experimental configuration in this run of the ADMX haloscope was similar to that reported in detail in Ref.~\cite{Khatiwada_2021} and updated in Ref~\cite{PhysRevLett.127.261803}. In previous runs, ADMX excluded both benchmark models for QCD axion dark matter over several hundred MHz~\cite{PhysRevLett.127.261803,PhysRevLett.120.151301,PhysRevLett.124.101303,PhysRevD.69.011101,PhysRevD.64.092003}. For the operations reported in this letter, the resonator system was outfitted with a different tuning rod, an improved noise calibration load, and improved cryogenic heat sinking, all with the aim of improving thermal and RF performance in the target frequency range.

\begin{figure*}[t]
\includegraphics[width=\textwidth]{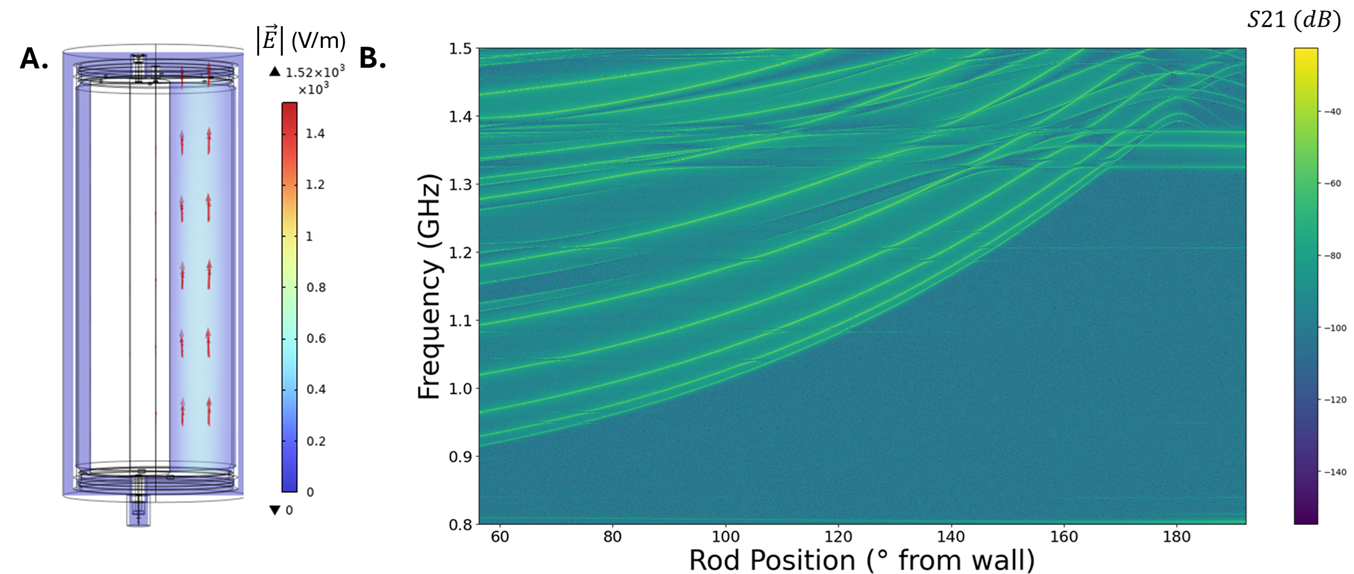}
\caption{\label{fig:cavity} A: The electric field map of the TM$_{010}$ mode simulated by COMSOL Multiphysics~\cite{comsol} where the direction is marked by the red arrows and the field magnitude by the color corresponding to the adjacent scale bar. B: A mode map of the cavity showing its resonant frequencies as the rod is moved. The heatmap denotes the $S_{21}$ transmission through the resonator, with bright lines indicating the resonant modes of the cavity. The mode used in the search is the fundamental $TM_{010}$ mode, which can tune from $900-1400$~MHz.}
\end{figure*}

The resonator used by ADMX in this experimental run is a right cylindrical copper plated stainless steel cavity (with length $101.4$~cm and diameter $41.9$~cm) tuned by a single $20.3$~cm copper tuning rod placed inside the 139-liter cavity, as shown in Fig.~\ref{fig:cavity}. The cavity was centered inside a solenoidal magnet. The magnet is energized by a constant current source delivering $225$~A and is stable to within $1$~mA. The magnetic field along the solenoid's longitudinal axis decreases from $7.8$~T at the central midplane to $5.3$~T at the top and bottom ends of the cavity. An off-center axle enabled rotation of the tuning rod from the wall of the cavity to near the center of the cavity, changing the resonant frequency. Across the frequency range of the experiment, the cavity unloaded quality factor was about 90,000. Two dipole antennas were inserted into the cavity for signal injection and readout. The first was a ``weakly coupled" antenna inserted into the bottom of the cavity. It is fixed to be weakly coupled to the cavity. The second was a ``strongly coupled" antenna, inserted into the top of the cavity. The coupling of the antenna, $\beta$, was controlled by a linear gearbox which adjusted the insertion depth of the antenna. The ability to vary the coupling ensured that the antenna was kept in an overcoupled state to optimize the scanning for axions across the full tuning range of the run~\cite{Kim_2020}. Both the tuning rod and antenna gearboxes are activated by room-temperature stepper motors via a long G10 fiberglass shaft to enable automated adjustment over the course of the entire run. 

When tuned to the frequency of the photon ($\nu_a=\frac{m_ac^2}{h}$) produced from the Primakoff effect, a persistent and coherent signal would develop within the cavity. The axion signal would follow an assumed Maxwell-Boltzmann distribution associated with the kinetic-energy distribution of dark matter within the local Milky Way halo~\cite{PhysRevD.42.3572}.The expected power coupled out of the cavity of an axion haloscope is~\cite{PhysRevLett.51.1415}
%
\begin{align*}\label{eqn:axion}
&P_{\mathrm{axion}}=1.8\cdot 10^{-23}~\text{W} \left(\frac{V}{106~\ell}\right)\left(\frac{B}{7.6~\text{T}}\right)^2\left(\frac{C}{0.4}\right)\\&\left(\frac{g_\gamma}{0.36}\right)^2
\left(\frac{\rho_a}{0.45~\text{GeV/cc}}\right)\left(\frac{\nu_a}{800~\text{MHz}}\right)\left(\frac{Q_L}{45,000}\right)\\&\left(\frac{\beta}{1+\beta}\right)\left(\frac{1}{1+(2\delta\nu_a/\Delta\nu_c)^2}\right),\numberthis
\end{align*}
%
where $V$ is the volume of empty space in the cavity, $B$ is the magnitude of the external magnetic field, $C$ is the cavity form factor (the overlap between the cavity resonant mode and the magnetic field), $g_\gamma$ is the model-dependent numerical constant ($-0.97$ the KSVZ model and $0.36$ for the DFSZ model) which determines the axion-photon coupling, $\rho_a$ is the expected dark matter density in the cavity, $Q_L$ is the loaded quality factor of the cavity, which is related to the unloaded quality factor by the equation $Q_0=Q_L(1+\beta)$, $\beta$ is the coupling coefficient of the antenna to the cavity mode, $\delta\nu_a$ is the difference between the cavity resonance and axion signal frequency, and $\Delta\nu_c=\frac{\nu_a}{Q_L}$ is the cavity linewidth. Here, the equation of the power has been normalized to reflect typical experimental parameters observed in ADMX. For axion haloscopes, the axion coupling to photons is related to the model-dependent constant by the equation

\begin{equation}
    g_{a\gamma\gamma}=\frac{\alpha g_\gamma}{\pi f_a}
\end{equation}
where $\alpha$ is the fine structure constant and $f_a$ is the axion decay constant.

The power from the $TM_{010}$ mode of the cavity is coupled out of the cavity by the strongly coupled antenna and transmitted into an ultra-low-noise microwave receiver for signal amplification and readout. Signals from the cavity are amplified with a first-stage Josephson parametric amplifier (JPA) \cite{9134828}, then a second-stage broadband heterodyne field effect transistor (HFET) amplifier \cite{bishop2006pattern}, before being transmitted to a room temperature receiver for further amplification and readout by a digitizer. The JPA used was a current-pumped design in which the resonant frequency of the JPA can be tuned by a bias current~\cite{Siddiqi_2004}. The bias current is used to match the JPA resonant frequency to the cavity resonant frequency.  The JPA is cooled to $105$~mK and typically provided 20~dB of power gain over a 20~MHz instantaneous bandwidth.  The HFET amplifier ($\rm LNF\text{--}LNC0.6\_2A$) provided an additional 33 dB power gain and is thermalized to 5 K~\cite{LNFLNC026A}. Circulators are placed at the inputs of the JPA and HFET for isolation. A cryogenic switch at the output of the antenna enables switching of the receiver input between the cavity and a ``heated load" noise source for calibrating the noise power of the experiment in situ. The JPA, circulators, and cryogenic switch are all sensitive to magnetic fields and are thus kept in a field-free region generated by a bucking coil magnet located about $1$~m above the main magnet. The fringe fields from the main solenoidal magnet are canceled by the bucking coil magnet to below $15$~mT and the JPA is additionally protected with passive $\mu$-metal shielding~\cite{Khatiwada_2021}. 

ADMX utilizes a He$^3$/He$^4$ dilution refrigerator to cool the system to minimize the thermal noise. The coldest part of the dilution refrigerator, the mixing chamber, achieved a temperature of 90~mK. This is mounted to the top of the cavity. The cavity was cooled to 140~mK and a copper cold finger mounted to the top of the cavity  conductively cooled the JPA and other field-sensitive electronics to 105~mK.





The operations covered in this letter began in December 2023 and continued until November 2024, the longest continuous data taking operation of the experiment. Over the course of the run, the data collection was automated and controlled using a series of control scripts. The cadence of the data taking procedure was as follows. First, a transmission measurement was performed by sending a swept signal through the cavity's weakly-coupled antenna and measuring the proportion of signal transmitted through the cavity to the critically coupled antenna. The transmission measurements were used to measure the resonant frequency and quality factor of the cavity. Following that, a reflection measurement was made, in which a swept signal was transmitted through a bypass line and reflected off the strongly coupled antenna of the cavity. The proportion of power reflected off the antenna was measured to determine the coupling coefficient of the antenna. The swept signal was then shut off and the power from the cavity was sampled for 100 seconds at a 100 kHz rate to search for potential axion signals. During the digitization process, a synthetic axion generator (SAG) system could transmit synthetic axions into the cavity at set frequencies through the weakly coupled port of the cavity. Unblinded synthetic axions were injected to characterize the receiver response, and blinded signals were used to test the robustness of our candidate identification methodology. Afterwards, the resonant frequency of the cavity was tuned and the data-taking process was repeated. 

The noise performance of the receiver was re-optimized in a JPA signal-to-noise ratio improvement (SNRI) measurement every 5-10 cycles. In this procedure, detailed in Ref.~\cite{khatiwada2020axion}, the JPA bias current and pump tone power were swept across multiple settings and the relative change in gain and noise power of the receiver were measured. The SNRI was then calculated as~\cite{doi:10.1063/1.1770483}

\begin{equation}
    \rm SNRI=\frac{G_{\rm on}}{G_{\rm off}}\cdot\frac{P_{\rm off}}{P_{\rm on}},
\end{equation}
where $G_{\rm on}$, $G_{\rm off}$, $P_{\rm off}$, $P_{\rm on}$ are the gain and noise power from the receiver with the JPA on and off, respectively. The system noise temperature of the experiment was calculated at 

\begin{equation}
    \rm T_{sys}=T_{HFET}/SNRI,
    \label{Eqn:Tsys}
\end{equation}
where $T_{\rm HFET}$ is the noise temperature of the receiver elements downstream of the JPA, mostly dominated by the second stage HFET amplifier~\cite{receivernoiseaxionhaloscopes}. The method for measuring $T_{\rm HFET}$ is discussed below. 

To calibrate the power coming from the experiment and the effective system noise temperature, two methods were used, enabling cross-verification of the system noise temperature. For both methods, the receiver was switched to the heated load. The first method calibrated the system noise temperature via a Y-factor measurement with the pump tone of the JPA turned off, such that the JPA was turned off and behaved as an ideal reflector. In this configuration, the noise contribution from the receiver was dominated by the second stage HFET amplifier. As such, the result of this fit was denoted as the effective HFET noise temperature, $T_{\rm HFET}/\alpha_{\rm eff}$. Here, $\alpha_{\rm eff}$ is a measure of the transmissivity of the signal path between the strongly coupled antenna of the cavity and the HFET, and the ``effective'' subscript is due to small noise contributions from circulators. The second method calibrated the system noise temperature via a Y-factor measurement with the JPA pump tone turned on so that the JPA behaved as an amplifier. In this configuration, the noise contribution was dominated by the JPA because all downstream noise was suppressed by the JPA gain. Therefore, the result of this fit is denoted as the effective JPA noise, $T_{\rm JPA,eff}$, where the ``effective'' subscript is included for the same reason as it was for the HFET fit result.  



This was the first data taking run with ADMX where the JPA on Y-factor measurement was included for calibration. This was enabled by upgrades to our calibrated noise source which improved the thermal isolation between the noise source and the mixing chamber. The additional Y-factor measurement including the JPA allowed us to check that our models for calculating the overall system noise temperature were consistent. $T_{\rm sys}$ was then calculated using Eqn.~\ref{Eqn:Tsys}.

\begin{figure}
\includegraphics[width=0.5\textwidth]{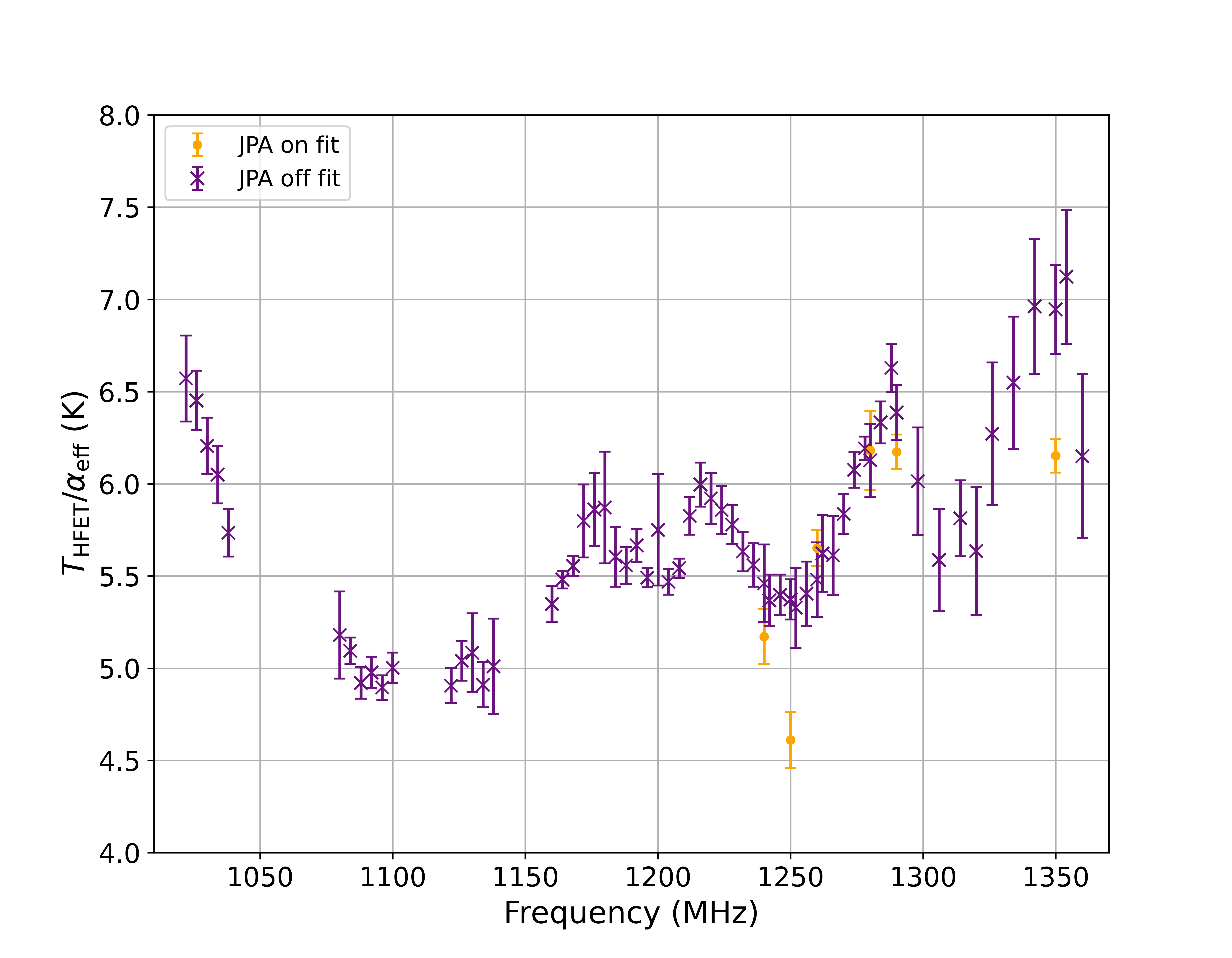}
\caption{\label{fig:noise} The effective HFET noise, $T_{\rm HFET}/\alpha_{\rm eff}$, over a range of frequencies. Gaps in the frequency data correspond to frequency regions that were not probed for axions. We measured $T_{\rm HFET}/\alpha_{\rm eff}$ using two different methods. The first method uses the direct fit of $T_{\rm HFET}/\alpha_{\rm eff}$  from a Y-factor measurement done with the JPA pump tone powered off (labeled JPA off fit). The second method uses the result from a Y-factor measurement done with the JPA pump tone powered on, which measures the JPA effective noise ($T_{\rm JPA,eff}$). We then use the SNRI and the models described in \cite{receivernoiseaxionhaloscopes} to calculate what the corresponding $T_{\rm HFET}/\alpha_{\rm eff}$ is for each value of $T_{\rm JPA,eff}$ (labeled JPA on fit). The error bars shown are statistical and systematic errors. Larger error bars are due to systematic errors associated with the fit to the Y-factor measurement.}
\end{figure}

A JPA on Y-factor measurement was done at 6 frequencies (1240, 1250, 1260, 1280, 1290, 1350 MHz). The JPA off Y-factor measurement was done at the same frequencies for comparison purposes, as well as many others to increase the total frequency coverage of our noise calibrations. Fig.~\ref{fig:noise} shows the values of $T_{\rm HFET}/\alpha_{\rm eff}$ at each of the 6 frequencies using both methods. The points labeled ``JPA off fit" use the direct fit result from the JPA off Y-factor measurement. Meanwhile, the points labeled ``JPA on fit" points combine the JPA on Y-factor fit results with the SNRI (using methods described in Ref.~\cite{receivernoiseaxionhaloscopes}) to calculate the corresponding value of $T_{\rm HFET}/\alpha_{\rm eff}$. For the frequencies where we took both types of measurements, we can also use the fit results to calculate the system noise temperature (the quantity we ultimately want to know) in two different ways. On average, there was a 5\% difference between the system noise temperature calculated using the two different methods, which is propagated to the systematic uncertainties in $T_{\rm sys}$ in the analysis. This difference indicated that the two methods were consistent at the few-percent-level. In the primary data analysis, we used the JPA off method because the calibration could be done at multiple frequencies at a time due to the broadband nature of the HFET amplifier, enabling a dense set of $T_{\rm sys}$ over the entire frequency range. Overall, for this data taking run the mean value of $T_{\rm sys}$ was 0.59 +/- 0.31~K. 

The individual spectra collected during data taking were combined into a ``grand spectrum" to optimize the signal-to-noise ratio of a potential axion signal using the method described in Ref.~\cite{PhysRevD.103.032002,Brubaker:2017rna}. Throughout the course of data-taking operations, axion candidates were defined as regions in the grand spectrum in which we could not exclude a KSVZ axion.

After candidates had been identified during data taking, each candidate frequency was subject to a veto procedure to evaluate whether or not the signal was consistent with that of an axion. The first step was to implement a rescan procedure in which candidate frequencies were scanned again to identify whether the excess power at that frequency was persistent. The rescan procedure was performed twice; the first rescan occurred with no changes to the data acquisition system and the second rescan occurred after turning off all blinded synthetic injections.

Any candidates that persisted through rescans were subject to additional checks. First, the signal power was evaluated as a function of ${\delta}\nu_a$, its distance from the resonant frequency. An axion-like signal would maximize on resonance, as seen in equation~\ref{eqn:axion} and follow the expected Lorentzian shape as a function of frequency. Candidates that exhibited power independent of the cavity resonant frequency are considered to be radio frequency interference (RFI) introduced downstream of the cavity and were eliminated. Furthermore, candidates that could be detected with a room temperature antenna and spectrum analyzer were considered to be RFI and eliminated from the list.

\begin{figure*}[t]
\includegraphics[width=1.0\textwidth]{figures/Run1D_limit_with_inset_9_2_25.png}
\caption{\label{fig:limit} 
A global limit plot putting this work (shown in purple) in context with other experiments, with an inset zooming in on this work's 90\% C.L. upper limits on $g_{a\gamma\gamma}$ (as well as limits from Ref.~\cite{CAPP_prx} due to overlapping coverage). The dark matter density is assumed to be 0.45 GeV/$\mathrm{cm^3}$. Gaps in the limits are due to mode crossings, regions where axion search mode of the cavity intersected other static weakly tuning modes. KSVZ and DFSZ sensitivities are shown as dashed lines. Previous limits set by ADMX are shown in teal \cite{PhysRevLett.104.041301,PhysRevLett.120.151301,PhysRevLett.124.101303,PhysRevLett.127.261803,bartram2024axiondarkmatterexperiment,10.1063/5.0122907,PhysRevLett.121.261302}. Limits from other experiments depicted include those set by University of Florida (UF)~\cite{uf_limit_1990}, Rochester-Brookhaven-Florida (RBF)~\cite{Wuensch:1989sa}, Center for Axion and Precision Physics (CAPP)~\cite{Lee:2020cfj,Jeong:2020cwz,CAPP:2020utb,Lee:2022mnc,Kim:2022hmg,Yi:2022fmn,Yang:2023yry,Kim:2023vpo,CAPP_prx}, Haloscope At Yale Sensitive To Axion Cold dark matter (HAYSTAC)~\cite{HAYSTAC:2018rwy,HAYSTAC:2020kwv,HAYSTAC:2023cam}, Grenoble Axion Haloscope project (GrAHal)~\cite{Grenet:2021vbb}, Oscillating Resonant Group AxioN experiment (ORGAN)~\cite{McAllister:2017lkb,Quiskamp:2022pks,Quiskamp:2023ehr}, MAgnetized Disc and Mirror Axion eXperiment (MADMAX)~\cite{c749-419q}, QUaerere AXions experiment (QUAX)~\cite{Alesini:2019ajt,Alesini:2020vny,Alesini:2022lnp,QUAX:2023gop,QUAX:2024fut}, Relic Axion Dark Matter Exploratory Setup (RADES)~\cite{CAST:2020rlf}, Taiwan Axion Search Experiment with Haloscope (TASEH)~\cite{TASEH:2022vvu}, CAST-CAPP~\cite{Adair:2022rtw}, and CERN Axion Solar Telescope (CAST)~\cite{cast2024}.}
\end{figure*}

Signals that passed the RFI checks were then rescanned using the TM$_{011}$ mode of the cavity. The TM$_{011}$ mode should have almost no coupling to the axion due to its reduced form factor, and thus any signals observed on the TM$_{011}$ mode would be due to external sources. Candidates that were not eliminated would then be tested with a magnet ramp procedure in which the power of the candidate is studied as a function of the external magnetic field. According to Eq.~\ref{eqn:axion}, the power of a true axion signal would follow the square of the magnetic field 

Over the course of the run, a number of candidate axion signals were observed. Table \ref{Tab:Candidates} lists the candidates that passed the initial persistence check, as well as the results from the candidate veto procedure. Eight were confirmed as blind signals injected by the SAG system after the second rescan. One was identified as RFI with the ambient check. One extra blinded SAG was identified and eliminated with the TM$_{011}$ scan. After all candidate axions were ruled out, limits were placed on the axion-photon coupling  across the covered mass range. At the conclusion of data-taking, the cavity was removed from the magnet bore. 

Over the course of the run, ADMX achieved better-than KSVZ sensitivity across the explored frequency range. The sensitivity accounted for the signal efficiency (90\%) due to the Savitzky–Golay filter parameters used in this run. The signal efficiency was evaluated by injecting software generated synthetic axion signals and evaluating the signal strength through the analysis chain.  
\begin{table}[htb!]
\caption{\label{Tab:Candidates} Overview of persistent candidates. ``SAG'' marked the candidates eliminated by turning off the SAG system (a blinded SAG wouldn't be turned off at this stage). ``RFI'' flagged the candidate eliminated by the ambient signal check or the enhancement-on-resonance check. ``TM$_{011}$'' flagged the candidate eliminated by the scan with the TM$_{011}$ mode which was a blinded SAG.}
\begin{ruledtabular}
\begin{tabular}{c c} 
    Frequency [Hz]& Veto Reason   \\\hline
    1,247,550,000  & SAG  \\
    1,247,320,000  & SAG  \\ 
    1,245,610,000  & SAG  \\
    1,193,709,600  &  SAG \\
    1,192,160,000 & SAG  \\ 
    1,138,850,000 & SAG \\
    1,131,860,500 & SAG \\
    1,128,120,000 & SAG\\
    1,130,613,500 & TM$_{011}$\\
    1,089,999,500 & RFI\\
\end{tabular}
\end{ruledtabular}
\end{table}


To place limits on $g_{a\gamma\gamma}$, the axion's coupling to photons, the signal-to-noise ratio (SNR) in each bin within the grand spectrum was compared to a Maxwell-Boltzmann axion lineshape model using a maximum likelihood weighting.  The uncertainty on power was then de-weighted with the systematic uncertainties shown in Table~\ref{tab:uncer_summary}. 

The uncertainties are similar to those from  previous ADMX runs~\cite{PhysRevLett.127.261803}, with an additional Lorentzian lineshape uncertainty. This additional contribution is due to a kHz-level drift of the cavity resonance during the 100~s axion search digitization observed in this run. The uncertainty due to this drift is listed in Table~\ref{tab:uncer_summary}. The uncertainties in the quality factor, $Q_L$, and coupling $\beta$ are determined from uncertainties in the fit to the transmission and reflection measurements, respectively. The uncertainty in $T_{sys}$ is determined from statistical and systematic uncertainties in the fit of the Y-factor measurement, which are discussed in further detail in ~\cite{PhysRevD.111.092012}. The uncertainty in $B^2VC$ is determined by variations in the magnetic field and form factor, as determined by simulation, when incorporating in the engineering tolerances of the resonator.

The confidence level of the exclusion was determined using a truncated normalized cumulative distribution function given by,

\begin{equation}
    \mathrm{C.L.}=\frac{\left[1-\mathrm{erf}\left(\frac{X-s}{\sqrt{2}}\right)\right]-\left[1-\mathrm{erf}\left(\frac{X}{\sqrt{2}}\right)\right]}{1-\frac{1}{2}\left[1-\mathrm{erf}\left(\frac{X}{\sqrt{2}}\right)\right]}.
\end{equation}
where $X$ is upper limit on the signal power and $s$ is the measured signal power. The distribution was truncated to not include unphysical negative couplings in accordance with the Feldman Cousins methodology~\cite{FeldmanCousins}.  From this analysis we can compute the exclusion limit for $g_{a\gamma\gamma}$ at 90\% confidence level as a function of frequency assuming a local dark matter density of $0.45$~GeV/cm$^3$ for the Maxwell-Boltzmann model.The limit set for this run is shown in Fig.~\ref{fig:limit}.  A global limit plot including the limit from this run in context with previously set limits is also shown in Fig.~\ref{fig:limit}. 

\begin{table}[htb]
\caption{\label{tab:uncer_summary} Summary of the averaged fractional percentange uncertainties associated with SNR.}
\begin{ruledtabular}
\centering
\begin{tabular}{lc}
    Source & Percentage Uncertainty on SNR\\
   \colrule
   $Q_L$ &  0.6 \%\\
   $\beta/(1+\beta)$ & 0.3 \% \\
   $T_{\rm sys}$ & 9.9\%\\
   $B^2 VC$ & 5\%\\
   Lorentzian shape & ~~1.5 \%\\
   Signal efficiency & 3\%\\
   Total & 11.4\%\\
\end{tabular}
\end{ruledtabular}
\end{table}

In summary, ADMX probed for QCD axions across the $1.10-1.31$~GHz frequency range during this reported data run. These results cover a previously unexplored region of parameter space and while an axion-like signal was not detected, new limits were placed on the axion-photon coupling for axion dark matter. 
These limits exclude KSVZ axions even at fractional dark matter densities. 

We were not able to achieve sensitivity to the DFSZ axion while maintaining a satisfactory scan rate during this run. ADMX is currently undergoing upgrades including an improved tuning system and thermal design. These upgrades will reduce the temperature of the cavity and RF electronics and improve the stability of experimental operations. In coming runs, ADMX will expand the covered parameter space probing downwards and out to probe for both the KSVZ and DFSZ axion to $1$~GHz.





This work was supported by the U.S. Department of Energy through Grants No. DE-SC0009800, No. DESC0009723, No. DE-SC0010296, No. DE-SC0010280, No. DE-SC0011665, No. DEFG02-97ER41029, No. DEFG02-96ER40956, No. DEAC52-07NA27344, No. DEC03-76SF00098, No. DE-SC-0022148 and No. DE-SC0017987. Fermilab is a U.S. Department of Energy, Office of Science, HEP User Facility. Fermilab is managed by Fermi Research Alliance, LLC (FRA), acting under Contract No. DE-AC02-07CH11359. CRB is supported by DOE Office of Science, High Energy Physics, Early Career Award (FWP 77794 at PNNL). Chelsea Bartram acknowledges support from
the Panofsky Fellowship at SLAC. John Clarke acknowledges support from the U.S. Department of Energy, Office of Science, National Quantum Information Science Research Centers. Additional support was provided by the Heising-Simons Foundation and by the Lawrence Livermore National Laboratory and Pacific Northwest National Laboratory LDRD offices. The Sheffield group acknowledges support from the UK Science and Technology Facilities Council (STFC) under grants ST/ T006811/1 and ST/X005879/1. UWA and Swinburne participation is funded by the ARC Centre of Excellence for Engineered Quantum Systems and CE170100009, Dark Matter Particle Physics, CE200100008. The corresponding author is supported by JSPS Overseas Research Fellowships No. 202060305. LLNL Release No. LLNL-JRNL-2004048. LANL Release No. LA-UR-25-2317.

\bibliography{references}

\end{document}